\documentclass[aps,prl,preprint]{revtex4}
\usepackage{epsfig}

\begin{document}

\title{EXACT AND APPROXIMATE SOLUTIONS FOR THE DILUTE ISING MODEL}
\bigskip

\author{Maurizio Serva}  
\affiliation{Dipartimento di Matematica,
Universit\`a dell'Aquila,
I-67010 L'Aquila, Italy}

\bigskip
\bigskip
\bigskip

\date{\today}

\begin{abstract}

The ground state energy and entropy of the dilute mean field
Ising model is computed exactly by a single order parameter.
An analogous exact solution is obtained in presence of a
magnetic field with random locations.
Results allow for a complete understanding of the geography
of the associated random graph.
In particular we give the size of the giant component (continent)
and the number of isolated clusters of connected spins of all given 
size (islands).
We also compute the average number of bonds per spin in the continent
and in the islands.
Then, we tackle the problem of computing the free energy
of the dilute Ising model at strictly positive temperature.
We are able to find out the exact solution in the paramagnetic region
and exactly determine the phase transition line.
In the ferromagnetic region we provide a
solution in terms of an expansion with respect to 
a second parameter
which can be made as accurate as necessary.
All results are reached in the replica 
frame by a strategy which is not based on multi-overlaps.


\end{abstract}

\maketitle

\section{1 - Introduction}

The study of the dilute mean field 
Ising model can be tackled using replica approach
which has been successfully applied to many disordered systems, 
the most celebrated being the SK model~\cite{SK,MPV}. 
Nevertheless, the method encounters serious difficulties when applied 
to many other systems because of a proliferation of replica order 
parameters as, for example, the multi-overlaps in dilute models~\cite{FL,GT,HAS}. 

In this paper we use a strategy proposed by Monasson~\cite{M},
which we already applied to the dilute Ising model~\cite{Se},
for which we computed the exact entropy and energy in the vanishing temperature case.
The Monasson strategy is very general and shows how
$<$$Z^n$$>$ ($Z$ is the partition function of
a generic spin model with disorder) can be expressed in terms of a maximum
with respect to the possible values of $2^n$ positive densities $x(\sigma)$
(where  $\sigma = (\sigma^1,\sigma^2,......,\sigma^n )$)
with the constraint $\sum_{\sigma} x(\sigma)=1$.

In next section, following~\cite{Se}, we derive the free energy
of the dilute Ising model in terms of the densities $x(\sigma)$
for a generic temperature and also in presence of
an external magnetic field with random locations.
In section 3, we consider the zero temperature case, in absence of magnetic field
and we show that the model is formally equivalent to a $2^n$ Potts model.
This observation allows to express the $x(\sigma)$ in terms of a single order
parameter and the exact solution is easily obtained.
The results coincide with those found in \cite{DSG}
where the approach is not based on replicas.
We also show that the exactly computed internal energy and entropy allows
to draw conclusions about the geography of the associated random graph.
In particular we determine the size of the giant component (continent),
the average size of the isolated clusters of connected spins (islands) and
the average number of bonds per spin in the continent and in the islands.
In section 4 we compute exactly the
ground state energy and entropy in presence of a randomly located magnetic field.
This allows for a more complete description
of the geography, in particular, the number of isolated clusters of 
connected spins of all given  size (islands) is determined.
In section 5 we extend the scope by considering the same model
in the case of a strictly positive temperature, where a second order 
parameter is needed.
We are able to find out the exact solution in the paramagnetic region
and exactly determine the phase transition line.
In the ferromagnetic region we provide a
solution in terms of an expansion with respect to the second order parameter
which can be made as accurate as necessary.
Conclusions and outlook are in the final section.

\section{2 - The dilute ferromagnet and the replica solution}

The dilute ferromagnetic Ising model is characterized by a number $M$
of order $N$ of non vanishing (and positive) bonds connecting pairs of spins. 
The partition function can be written as

\begin{equation}
Z= \sum_{\#} \exp \left( \beta \sum_{i>j} K_{ij} \sigma_i \sigma_j \right)
\label{partition1}
\end{equation}
where the sum $\sum_{\#}$ goes on the $2^N$ realizations
of the $N$ spin variables and the
$K_{ij}$ are quenched variables which take the value $1$ 
with probability $\frac{\gamma}{N}$ and  $0$ otherwise.
The dilution coefficient $\gamma$ may take any positive value
and the number $M$ of bonds is $\frac{\gamma}{2}  N \pm o(\sqrt{N})$.
This is at variance with the fully connected model
where the number of non vanishing bonds is $N(N-1)/2$ and it is the reason
why the dilute ferromagnetic has a richer behavior.

An alternative definition of the partition function
can be obtained by taking a number of bonds exactly equal to $M$
where $M$ can be both a deterministic number proportional to $N$ or a 
quenched random variable with average proportional to $N$ and fluctuations of 
order $\sqrt{N}$. In this context the pair of spin 
to be connected to a given bond is chosen randomly~\cite{GT, DSG}.
All these models and model (\ref{partition1}) are thermodynamically 
equivalent, i.e. the intensive thermodynamical quantities are the same.

Models with lesser dilution~ \cite{BC,BG} (number of bonds of order 
$ N^\epsilon $ with $1$$<$$\epsilon $$<$$2$ ) have been also considered, but in 
this case the thermodynamics is the same of the fully connected model.

The problem of evaluating (\ref{partition1}) via the replica method,
can be approached in many different ways. A possibility is to follow the
same path which has permitted the solution of the SK model using
multi-overlaps as replica order parameters.
This strategy leads to a proliferation of parameters and it is unable to
give exact answers~\cite{HAS}.
In~\cite{Se}, we considered a different strategy,
which is based, instead, on the densities $x(\sigma)$ which
are $2^n$ parameters since $\sigma = (\sigma^1,\sigma^2,......,\sigma^n )$
may assume $2^n$ possible values.

In~\cite{Se} we computed the partition function and, therefore,
the free energy $F$ obtaining

\begin{equation}
-\beta F =  \lim_{n \to 0}\frac{\Psi_n}{n}+
\frac{\gamma}{2}\log(\cosh(\beta)) 
\label{bf}
\end{equation}
where $\Psi_n$ is given by

\begin{equation}
\Psi_n = \max_{x}  \;  [ \, -\frac{\gamma}{2} + \frac{\gamma}{2}\sum_{\sigma,\tau} 
x(\sigma) x(\tau)
\prod_\alpha  (1+\tanh(\beta) 
\sigma^\alpha \tau^\alpha)
-\sum_{\sigma} x(\sigma) \log( x(\sigma)) \, ]
\label{psi}
\end{equation}
where the sum $\sum_{\sigma,\tau}$ goes on the $2^n$ possible 
values of the variable $\sigma$ and the the $2^n$ possible values of 
the variable $\tau$.
The maximum is taken with respect the $2^n$
densities $x(\sigma)$ with the constraints
$0 \le x(\sigma) \le 1$ and $\sum_\sigma x(\sigma) =1$.
The highly non trivial problem is the
maximization of the $x(\sigma)$ which,
in principle, can be found by a proper parametrization. 

The simplest case turns out to be the high temperature 
paramagnetic region where the maximum is reached when
$x(\sigma) = \frac{1}{2^n}$  for all $\sigma$, so that

\begin{equation}
-\beta F =  \log(2)+
\frac{\gamma}{2}\log(\cosh(\beta))
\label{bfhigh} 
\end{equation}

We will show in section 5
that the paramagnetic region is the high temperature region
given by $\gamma\tanh(\beta) \le 1$
(see also \cite{GT, DSG}). 
\smallskip

In~\cite{Se} we also considered the dilute ferromagnetic system in a magnetic field 
with random locations.
The partition function of this model is

\begin{equation}
Z= \sum_{\#} \exp \left( \beta \sum_{i>j} K_{ij} \sigma_i \sigma_j 
+ \beta \sum_{i} h_{i} \sigma_i  \right)
\label{partition2}
\end{equation}
where $K_{ij}$ are the previously defined quenched variables
and the $h_i$  take the positive value $h$ with probability $\delta$
and $0$ otherwise.

In this case we get
\begin{equation}
-\beta F =  \lim_{n \to 0}\frac{\Psi_n(\delta)}{n}+
\frac{\gamma}{2}\log(\cosh(\beta)) 
+ \delta \log(\cosh(\beta h)) 
\label{bfd}
\end{equation}
where $\Psi_n(\delta)$ is obtained by maximizing
an expression which is the same of that in (\ref{psi}) plus the extra term

\begin{equation}
 \sum_{\sigma} x(\sigma) 
\log \left( 1-\delta +\delta \prod_\alpha  (1+\tanh(\beta h)
\sigma^\alpha) \right)
\label{psid}
\end{equation}
where the sum $\sum_{\sigma}$ goes on the the $2^n$ possible values of the variable
$\sigma$. Obviously, $ \Psi_n(0)=\Psi_n$.

Before trying to compute $\Psi_n(\delta)$ in the general case,
we focus on the zero temperature case, which is exactly solvable
both in the vanishing (next section) and non vanishing 
(section 4) magnetic field case. In both cases many conclusions may be
drown concerning the structure of the random graph,
both for what concerns the giant component (continent)
and the small islands of connected spins.

\section{3 - Zero temperature and zero magnetic field}

Let us start by considering the case of vanishing temperature and
vanishing magnetic field ($\delta=0)$.
In the vanishing temperature limit one has $\tanh(\beta)=1$ and, therefore, 
$\prod_\alpha  (1+\tanh(\beta) \sigma^\alpha \tau^\alpha)$ 
becomes $\prod_\alpha  (1+\sigma^\alpha \tau^\alpha)$.
This term equals $2^n$ when all  $\sigma^\alpha$ equals $\tau^\alpha$ and
vanishes otherwise. Therefore, expression (\ref{psi}) becomes

\begin{equation}
\Psi_n= \max_{x}  \;  [ \,
-\frac{\gamma}{2}+ \frac{\gamma}{2} 2^n \sum_{\sigma} 
x(\sigma)^2 -\sum_{\sigma} x(\sigma) \log( x(\sigma)) \, ]
\label{psi0}
\end{equation}
which is a standard $2^n$-components Potts model \cite{Wu}.
The solution is known and can be found assuming that $2^n-1$ 
quantities $x(\sigma)$ take the value $\frac{1-\theta}{2^n}$
and one takes the value $\frac{1+(2^n-1)\theta}{2^n}$.
The state with different value can be any of the possible $2^n$,
we assume that is the one with all $\sigma^\alpha=1$ for all $\alpha$.
We can write:

\begin{equation}
x(\sigma) = \frac{1-\theta}{2^n} + 
 \frac{\theta\prod_{\alpha=1}^n (1+\sigma^\alpha)}{2^n}
\label{xpott}
\end{equation}
which satisfy the constraint $\sum_{\sigma} x(\sigma) =1$.

In~\cite{Se} we used  this solution in order to find the exact expression
for the ground state internal energy $E$  

\begin{equation}
E = -\frac{\gamma}{2}
\label{e0}
\end{equation}
while the entropy $S$ was found to be

\begin{equation} 
S= \log(2) \, \max_{\theta}  \;  [ \, \frac{\gamma}{2} 
( \theta^2-1)+ (1-\theta) (1-\log(1-\theta)) \, ]
\label{s0}
\end{equation}
The maximum is reached in $\theta_c$ given by the equation

\begin{figure}
\vspace{.2in}
\centerline{\psfig{figure=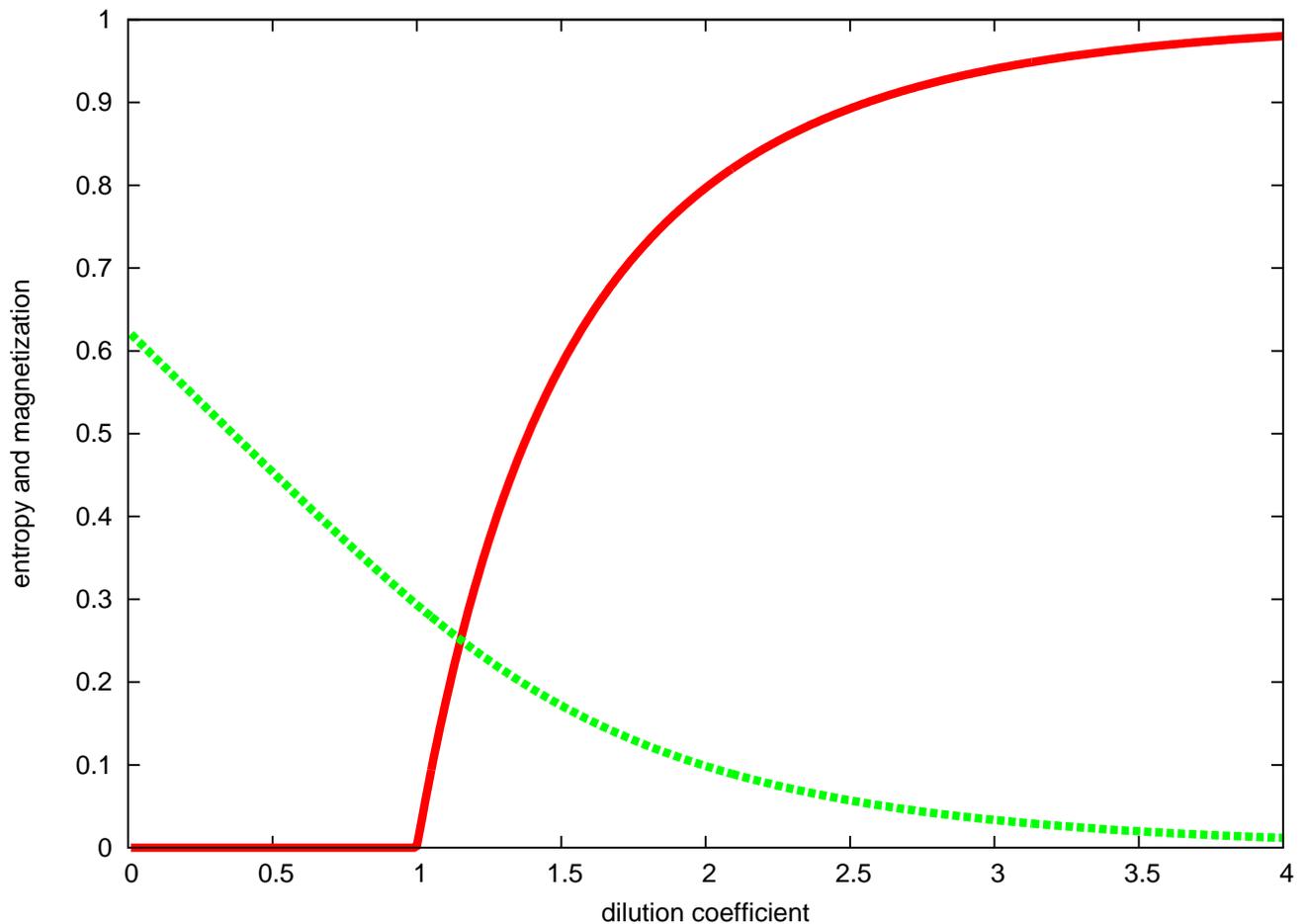,width=5.0truein,angle=270}}
\bigskip
\caption{
Entropy (dashed line) and order parameter $\theta_c$ 
(full line) as a function of
the dilution coefficient $\gamma$ at 0 temperature and 0 magnetic field.
The transition is at $\gamma=1$ where the first derivative of
$\theta_c$ and the third derivative of the entropy are discontinuous.}
\label{fig1}
\end{figure}

\begin{equation}
\exp(-\gamma\theta_c) = 1-\theta_c
\label{thetac}
\end{equation}
This equation has a single non negative solution $\theta_c=0 $ 
if  $\gamma \leq 1$
and one more non trivial positive solution if $\gamma > 1$ which corresponds
to the maximum.
Therefore, at 0 temperature, for  $\gamma \le 1$ the system is
in a paramagnetic phase ($\theta_c=0$) while for $\gamma >1$
is in a ferromagnetic phase  ($\theta_c>0$).
The parameter $\theta_c$ is the magnetization of the
system, but we will see that it has a simple interpretation
in terms of underling random graph and that its discontinuity  
corresponds to the percolation transition
generated by the ferromagnetic links.

Using equation (\ref{thetac}) we can rewrite the entropy (\ref{s0}) as

\begin{equation} 
S=\log(2)[(1-\frac{\gamma}{2}) 
+ (\gamma-1)\theta_c -\frac{\gamma}{2}\theta_c^2 ]
\label{s02}
\end{equation}
The entropy $S$ and the order parameter
$\theta_c$ are plotted in Fig. 1 as a function of
the dilution coefficient $\gamma$.
At the transition value $\gamma=1$, the first derivative of
$\theta_c$ and the third derivative of the entropy are discontinuous.

Indeed, these results have an interesting interpretation in terms of 
random graph geography. In fact, at zero temperature, the magnetization
is entirely due to the giant set (continent) of spins all connected,
while the small sets of a single or a few interconnected spins 
isolated from the others (islands)  cannot contribute to the magnetization.
Since all spins in the continent must be directed in the same direction
in order to minimize the ground state energy,
their magnetization must be 1 and, therefore,
their number must be $\theta_c N$.
This also means that the total number of spins in the islands 
must be $(1-\theta_c)N$. In particular,
when $\gamma \le 1$ there is no continent ($\theta_c = 0$) and all
the spin are in islands, on the contrary, all spins are in the continent 
only when $\gamma \to \infty$ (in this limit $\theta_c = 1$).

It is also possible to count the number of islands, in fact, any 
of the islands of connected spins can live only in two configurations:
all spins up or all spins down. Therefore, any island contribute by 
a $\log(2)/N$ to the (intensive) entropy. 
The continent also contributes by a single $\log(2)/N$
but this is irrelevant in the thermodynamical limit. Therefore, the number of 
island is simply $\frac{S}{\log(2)}N$.

Let us call $\alpha(l) N$ the number of islands of size 
$l$ (islands made by a number $l \ge 1$ of interconected 
spins which are isolated form all the others). 
Then, the total number of islands can be written as
$\sum_{l \ge 1} \alpha(l)\, N$. 
Furthermore, since an island of size $l$ contains by definitions $l$
spins, the  total number of spins in the islands
can be written as $\sum_{l \ge 1} \alpha(l)\, l \,N$ . 

We have seen that the contribution of any
island to the entropy is $\log(2)/N$, therefore, we can write

\begin{equation} 
S= \log(2) \sum_{l \ge 1} \alpha(l)
\end{equation}

Analogously, since the total number of spins in all the islands 
is $(1-\theta_c)N$, we can write

\begin{equation}
1-\theta_c=\sum_{l \ge 1} \alpha(l) \, l
\end{equation} 

We can also easily compute
the average size $<$$l$$>$ of the islands 
(the average number of spins per island) since it is given by the ratio between
the total number of spins in all the island $\sum_{l \ge 1} \alpha(l)\, l \,N$ 
divided by the number of islands $\sum_{l \ge 1} \alpha(l) \,N$. We obtain

\begin{equation} 
<l> = \frac{(1-\theta_c)\log(2)}{S}
\label{eta}
\end{equation}

\begin{figure}
\vspace{.2in}
\centerline{\psfig{figure=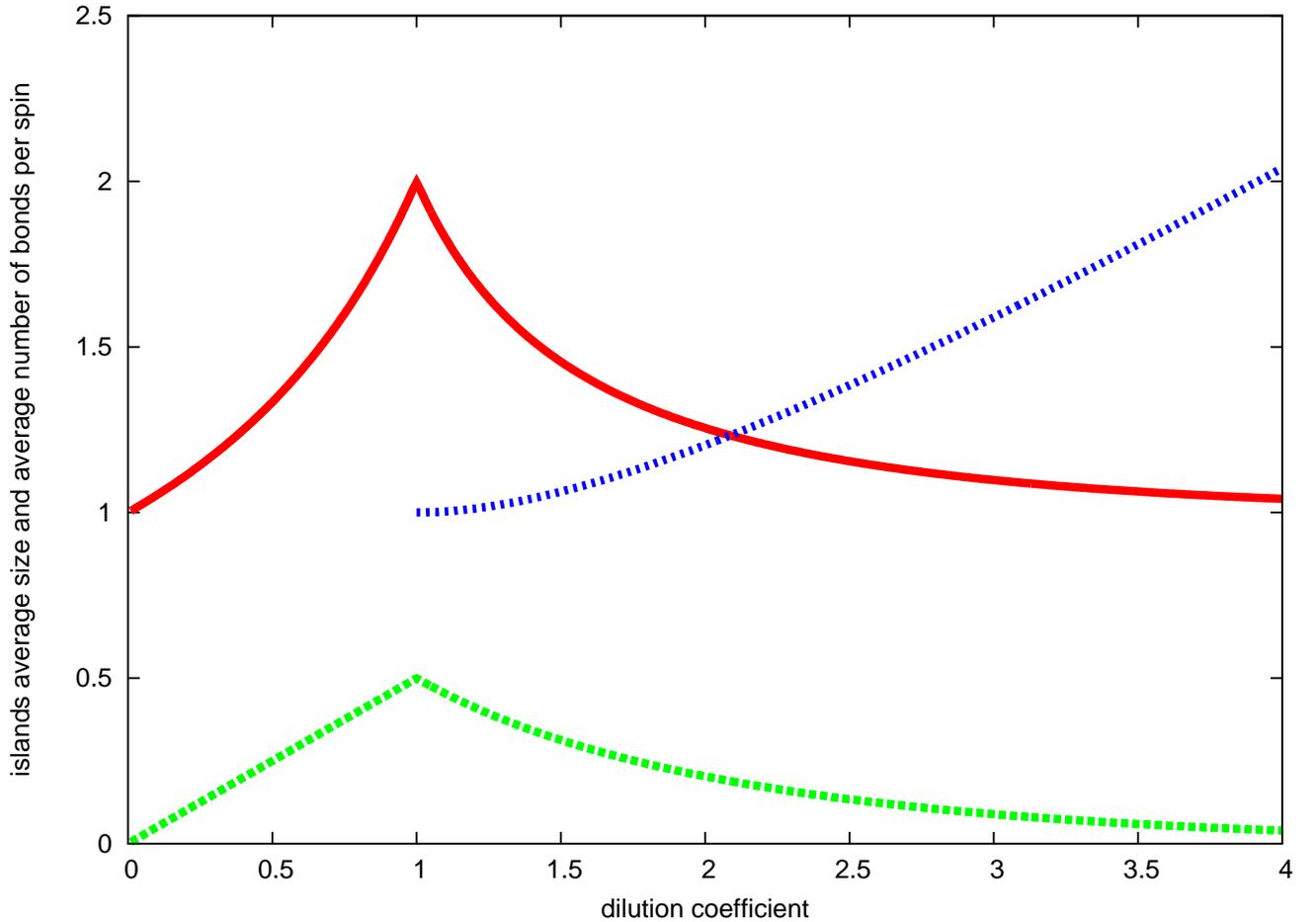,width=5.0truein,angle=270}}
\bigskip
\caption{
Average size $<$$l$$>$ of the islands (full line),
average number of bonds per spin in
the islands $\phi/(1-\theta_c)$ (dashed line) and average number of bonds 
per spin in the continent $(\gamma/2 -\phi)/\theta_c$ (dotted line).
The average number per spin in the continent is computed only when 
the continent exists ($\theta_c >0$), that's why the dotted 
line begins in $\gamma=1$. 
}
\label{fig2}
\end{figure}

The average size $<$$l$$>$ of the islands is plotted in Fig. 2, 
notice that it has a maximum 
at the transition $\gamma=1$ where it equals 2.
This means that the average dimension reaches a maximum when
the continent begins to exist and then decreases because larger
island are more  easily absorbed by the continent. 

Another simple consideration allows to compute
the number of bonds $\phi N$ in the islands. 
In fact, a moment of reflection shows that the number of islands forming
loops is irrelevant in the thermodynamical limit.
Therefore the spins of an island of size $l$ must be connected by $l-1$ bonds
which implies 

\begin{equation} 
\phi = \sum_{l \ge 1} \alpha(l) \, (l-1)
= 1-\theta_c - \frac{S}{\log(2)}
\label{phi}
\end{equation}

Since the total number of bonds is $\frac{\gamma}{2}N$,
we also have that number of bonds in the continent is $(\gamma/2 -\phi)N$.
Finally, we can compute the average number of bonds per spin in
the islands as $\phi/(1-\theta_c)$, while the average number of bonds 
per spin in the continent is $(\gamma/2 -\phi)/\theta_c$.
This two quantities are plotted in Fig. 2. The average number of bonds per spin in
the islands follows the average dimension of the island and it has a maximum in 
$\gamma=1$ where it is the value 0.5. This is coherent
with the fact that an island of size two as a single bond.
The average number of bonds per spin in
the continent is defined only when $\gamma > 1$ because this the region where the
continent exists. At $\gamma \to 1$ it has its minimal value 1
coherently with
the fact that the any of the spin must be 
connected at least to another spin, which is, in turn connected to a third one
and so on.

In next section we will consider again the case of vanishing temperature, but
in presence of a magnetic field. This will allow for
a better comprehension of the random graph geography.
In fact, we will be able to compute all the
$\alpha(l)$, for the moment we just remark that
the number $\alpha(1)N$ of islands of size $l$ (isolated spins)
is $N \exp(-\gamma) $.
This can be computed by considering that $\exp(-\gamma) $
is the thermodynamical limit of
$(1-\frac{\gamma}{N})^{N-1}$ which, in turn, is the probability that 
a spin has no positive bonds connecting to all the others.

\section{4 - Zero temperature and non zero magnetic field}

In the vanishing temperature limit
one has $\tanh(\beta)=\tanh(\beta h)=1$
and $\Psi_n(\delta)$ becomes

\begin{equation}
\Psi_n(\delta)=  \max_{x}  \;  [
-\frac{\gamma}{2}+ \frac{\gamma}{2} 2^n \sum_{\sigma} 
x(\sigma)^2 + 
\sum_{\sigma} x(\sigma) 
\log( 1-\delta +\delta \prod_\alpha  (1+\sigma^\alpha))
-\sum_{\sigma} x(\sigma) \log( x(\sigma)) \, ]
\label{gtpottm}
\end{equation}

\begin{figure}
\vspace{.2in}
\centerline{\psfig{figure=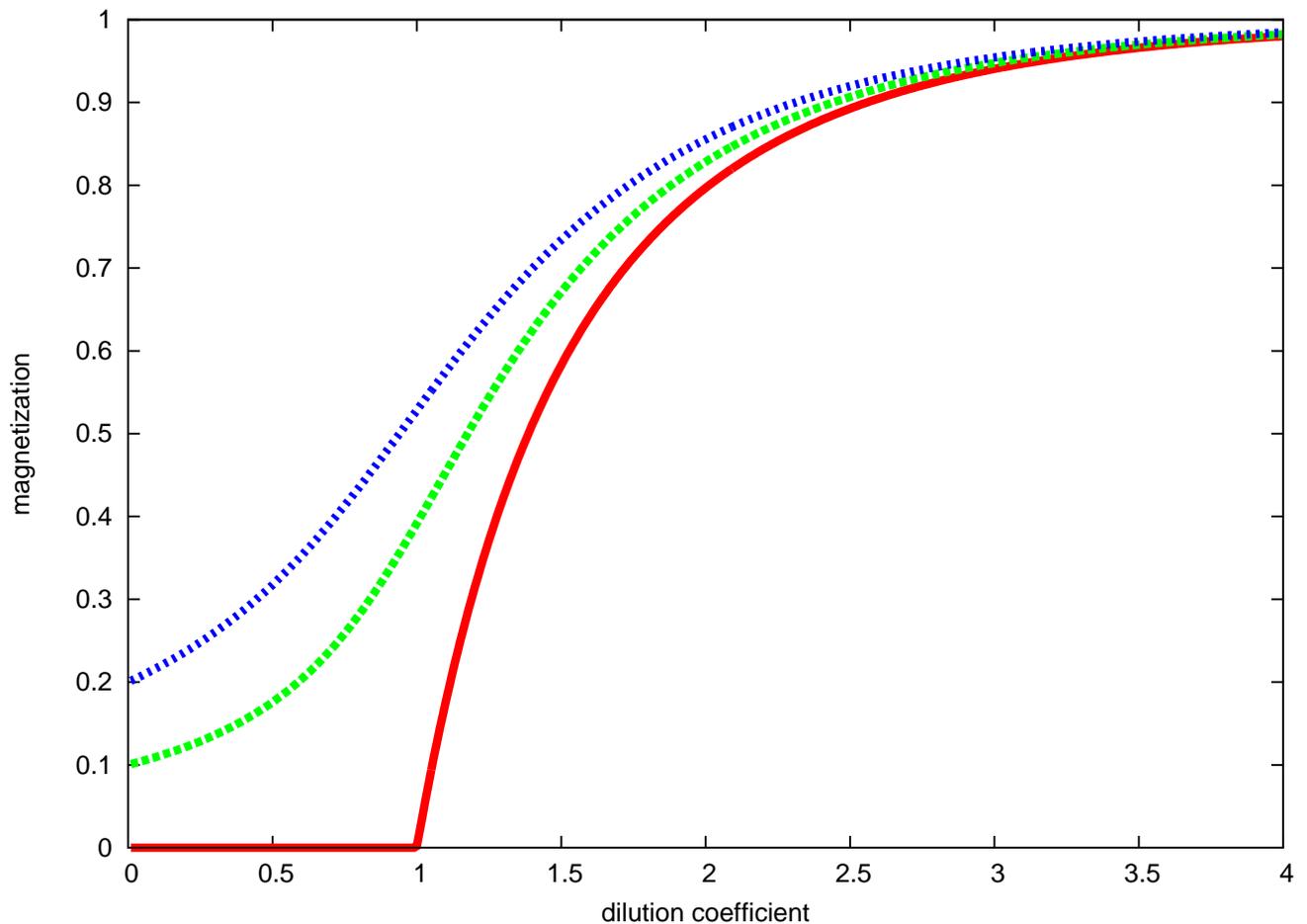,width=5.0truein,angle=270}}
\bigskip
\caption{
Magnetization $\theta_c(\delta)$ as a function of
the dilution coefficient $\gamma$ at three different values of 
the magnetic field concentration: $\delta=0$ (full line), 
$\delta=0.1$ (dashed line) and $\delta=0.2$ (dotted line). 
One can easily remark the absence of transition for
strictly positive values of $\delta$.
}
\label{fig3}
\end{figure}

This is again a Potts model whose solution can be found assuming  
that the $2^n$ densities $x(\sigma)$ 
have the form (\ref{xpott}). 

With some simple algebra~\cite{Se} we can compute the ground
state internal energy $E(\delta)$

\begin{equation}
E(\delta) = -\frac{\gamma}{2} -\delta h
\label{ed0}
\end{equation}
while the entropy $S(\delta)$ is

\begin{equation}
S(\delta)= \log(2) \,\max_{\theta}  \;  [ \,
\frac{\gamma}{2} 
( \theta^2-1)+ (1-\theta)\log(1-\delta) +
(1-\theta) (1-\log(1-\theta)) \, ]
\label{sd0}
\end{equation}

These two quantities obviously coincide
with those computed in the
previous section when $\delta=0$, i.e. $E(0)=E$ and $S(0)=S$. 
The maximum is reached in $\theta_c(\delta)$ given by the equation

\begin{equation}
(1-\delta) \exp(-\gamma\theta_c(\delta)) = 1-\theta_c(\delta)
\label{thetadc}
\end{equation}

At variance with the 0 magnetic field case,
this equation has a single non negative solution $\theta_c(\delta)$
for any value of $\gamma$
and, therefore, the transition disappears.
Also for this quantity one has by definition $\theta_c(0)=\theta_c$
where $\theta_c$ is the zero magnetic field
magnetization computed in previous section.
Using (\ref{thetadc}), the entropy $S(\delta)$ can be rewritten as

\begin{equation} 
S(\delta)=\log(2)[(1-\frac{\gamma}{2}) 
+ (\gamma-1)\theta_c(\delta) -\frac{\gamma}{2}\theta_c(\delta)^2 +
(1-\theta_c(\delta))\log(1-\delta)]
\label{sd02}
\end{equation}

In Fig. 3 we plot the magnetization $\theta_c(\delta)$ as a function of
the dilution coefficient $\gamma$ at three different values of 
the magnetic field concentration: $\delta=0$, $\delta=0.1$ and $\delta=0.2$. 
One can easily remark the absence of transition for
strictly positive values of $\delta$.

Finally we compute the entropy response $\sigma(\delta)$
to magnetic concentration $\delta$ as

\begin{equation}
\sigma(\delta)=-\frac{dS(\delta)}{d\delta}=
-\frac{\partial S(\delta)}{\partial \delta}
- \frac{\partial S(\delta)}{\partial \theta_c}
\frac{\partial \theta_c}{\partial \delta}
=-\frac{\partial S(\delta)}{\partial \delta}
\end{equation}
which gives

\begin{equation}
\sigma(\delta)=
\frac{1-\theta_c(\delta)}{1-\delta}
\label{sigma}
\end{equation}

In order to use the above results to obtain more
information concerning the geography of
the random graph, we preliminarily observe that
the difference between $\theta_c(\delta)$ and $\theta_c$ is only due to
the effect of the magnetic field on the islands while the magnetization
of the spins of the continents remains unchanged and equal to 1.
The same can be said for what concerns the entropy, in fact,
the contribution to the entropy
coming from the continent remains equal to zero.

It is easy to understand the effect of the magnetic field
on a given island: if none of the component spins get a 
positive magnetic field, the island can still
live in two configurations: all spins up
or all spins down. For this reason
its contribution to the entropy remains $\log(2)/N$ and the contribution 
to the magnetization remains equal to zero. 
But if one or more spins of the island get a positive
magnetic field, then all spins of the island must be up,
which implies that the contribution to the entropy vanishes and
the contribution to the magnetization
rises from 0 to $l/N$ where $l$ is the size of the island.

The interesting point is that while magnetization $\theta_c N$ has
a geometrical interpretation, simply representing
the number of spins in the continent, 
the magnetization $\theta_c(\delta)N $ is the sum of the number of spins
in the continent plus the number of all spins in the islands frozen by 
the magnetic field.
The entropy $S(\delta)N$, on the contrary, 
is simply the number of unfrozen islands
multiplied by $\log(N)$.

We notice that the probability that none of
the spins of an island of size $l$ get a magnetic field is
$(1-\delta)^l$ while the probability that at least one has a 
magnetic field is $1-(1-\delta)^l$.
The above arguments allow to say that the number of islands of size 
$l$ which are not frozen is $\alpha(l) (1-\delta)^l N$. 
Then, we are allowed to write

\begin{equation} 
S(\delta)= \log(2) \sum_{l \ge 1} \alpha(l)(1-\delta)^l
\label{sa}
\end{equation}
\begin{equation}
1-\theta_c(\delta) =\sum_{l \ge 1} \alpha(l)(1-\delta)^l \, l
\label{ta}
\end{equation}
which coincide with the analogous quantities in previous section
when $\delta=0$.
 
It is easy to verify that  (\ref{sa}) and (\ref{ta})  are coherent
with equation (\ref{sigma}) for the entropy response.
Furthermore, they give much more information,
in fact, expanding $1-\theta_c(\delta)$ 
for small values of $1-\delta$
we have that the coefficients of
the expansion are  $\alpha(l) \, l$.
As a consequence, we can have in principle all the 
density numbers $\alpha(l)$ of islands of size $l$.

In order to perform this expansion we can use iteratively 
equation (\ref{thetadc}) and we obtain 

\begin{equation} 
\begin{array}{ccc}
\alpha(1) = \exp(-\gamma) \\
\alpha(2) = \frac{1}{2} \gamma \exp(-2\gamma)  \\
\alpha(3) = \frac{1}{2} \gamma^2\exp(-3\gamma) \\
\alpha(4) = \frac{2}{3} \gamma^3\exp(-4\gamma) 
\end{array}
\end{equation}
while, in general, for any $l \ge 2$, one can deduce from  (\ref{thetadc})
the recursive formula

\begin{equation} 
\alpha(l) = \frac{1}{l} \, \exp(-\gamma)
\sum_{k=1}^{l-1} \frac{1}{k!} \,  \gamma^k
\sum_{\{l_i\}} \,
\prod_{i=1}^{k}\alpha(l_i) \,  l_i
\label{al}
\end{equation}
where the second sum goes on all the possible values of the $k$
strictly positive integer numbers
$l_i$ which satisfy the constraint 
$l_1 + l_2 + \cdot \cdot \cdot + l_k=l-1$.
This recursive formula gives 
\begin{equation} 
\alpha(l) = \frac{l^{\,l-2}}{l \, !} \,
\gamma^{l-1} \exp(-\gamma l)
\label{sal}
\end{equation}
indeed, we are not able to rigorously prove (\ref{sal}) from
(\ref{al}), nevertheless, we checked that it is exact for all $l$
between 2 and 10. Much more important, we were able to
numerically compute $\sum_{l \ge 1} \alpha(l) $ and 
 $\sum_{l \ge 1} \alpha(l) \,l$  and verify that they coincide, respectively,
with $S$ and $1-\theta_c$.
In particular $\sum_{l \ge 1} \alpha(l) \,l$ equals 1 for any
$\gamma \le 1$.

Notice that the expression $\alpha(1) = \exp(-\gamma)$
was already found in previous section using very simple arguments. 
Also notice that in the $\gamma \to 0$ limit all $\alpha(l)$
vanish except $\alpha(1)$ which goes to 1.
This is simply due to the fact that in this limit all spins are isolated
which means that only islands of size 1 exist.

The description of the geography of the random graph is now completed,
since we are able, for any value of
$\gamma$, to give both the dimension of the continent $\theta_c$
and the number $\alpha(l)$ of islands of any fixed size.
Some relevant intuition about this geography can be obtained 
considering the islands index $r(m)$ defined as
\begin{equation} 
r(m) = \frac{\sum_{l = 1}^m \alpha(l)}
{\sum_{l \ge 1} \alpha(l)}=
\frac{\log(2)}{S}
\sum_{l = 1}^m \alpha(l)
\label{rm}
\end{equation}
which is the number of islands of size $l \le m$ divided by 
the total number of islands. The index $r(m)$ goes to 1 for
large m. In Fig.4  we plot this quantity for
$m=1,2,3$ and $4$. The index approaches 1 for large values of $m$
but, for $m=4$ it is still about 0.94 when $\gamma=1$.
The lower values of $r(m)$ at $\gamma=1$ mean that the largest islands appear
around the transition.

\begin{figure}
\vspace{.2in}
\centerline{\psfig{figure=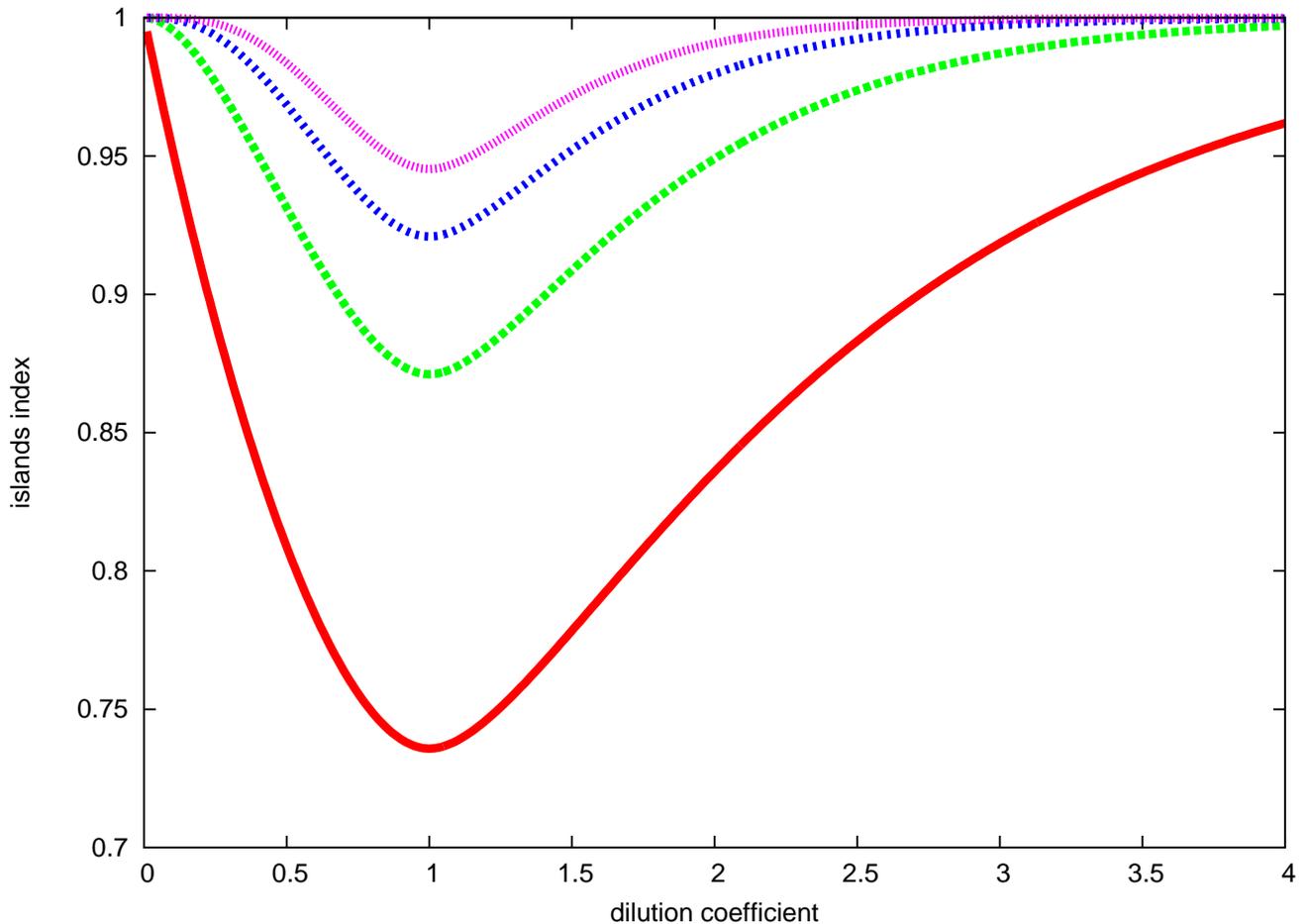,width=5.0truein,angle=270}}
\bigskip
\caption{ 
Number of islands $r(m)$ of size $l \le m$ divided by 
the total number of islands for $m=1,2,3$ and $4$
(from below). The index approaches 1 for larger values of $m$.
The lower values of $r(m)$ at $\gamma=1$ mean that the largest islands appear
around the transition.
}
\label{fig4}
\end{figure}

\section{5 - Positive temperature}

For a positive temperature, in absence of magnetic field, 
the free energy is given by (\ref{bf}) 
once the proper maximum in (\ref{psi}) is found.  
We have seen that in the case of vanishing temperature the exact 
solution can be found because the problem is equivalent to a Potts model,
on the contrary, for positive temperature, there is not
an analogous equivalence. 
Therefore, we have to find the correct parametrization of the $x(\sigma)$
making a decision motivated by the physics.
First of all we remark that
for positive temperature, the islands cannot contribute to the magnetization,
for the same reason why they do not contribute in the zero temperature case.
In fact, being isolated clusters of spins, they can be orientated
in both directions. So, we can already conclude that only the spins
in the continent may contribute to the magnetization. 
Let us assume that the magnetization of a spin in the continent is $\lambda$,
then it is natural to assume that

\begin{equation}
x(\sigma) = \frac{1-\theta_c}{2^n} + 
 \frac{\theta_c\prod_{\alpha=1}^n (1+\lambda \, \sigma^\alpha)}{2^n}
\label{xtemp}
\end{equation}
where $\theta_c$ is the continent size given by  (\ref{thetac}) 
and $0 \le \lambda \le 1$ is the parameter to be maximized.
Notice that the total magnetization is $m=\theta_c \lambda$, 
being simply the product of the magnetization of a spin in the continent
$\lambda$ and their density $\theta_c$. 
In fact,

\begin{equation}
m = \sum_{\sigma} x(\sigma) \sigma^\alpha =\theta_c \lambda
\label{mtemp}
\end{equation}
for any of the replicas, i.e. any $1\le \alpha \le n$.

It is clear, at this point, that the really crucial ansatz is
that the magnetization is a self-averaging quantity, otherwise we should 
chose an expression analogous to (\ref{xtemp}) 
but with a different $\lambda_\alpha$ correspondingly to any $\sigma^\alpha$.
This breaking of the symmetry would have implied a different magnetization
$m_\alpha$ for any possible  $1 \le \alpha \le n$.

With the choice (\ref{xtemp}) we have $\Psi_n = \max_{\lambda}  \Phi_n (\lambda)$,
where

\begin{equation}
\Phi_n (\lambda)= 
\frac{\gamma}{2} \theta_c^2 [(1+\lambda^2 \tanh(\beta))^n-1]
-\sum_{\sigma} x(\sigma) \log( x(\sigma)) 
\label{phitemp}
\end{equation} 
and where $x(\sigma)$ is given by (\ref{xtemp})
and the sum goes on all the $2^n$ possible realizations of $\sigma$.

In the zero temperature case ($\tanh(\beta)=1$ ) and vanishing $n$, we can easily check that
the maximum is reached in $\lambda=1$ and we recover the zero temperature solution
of section 2. This check can be performed by simply verifying that the derivative
with respect to $\lambda$ of $\Phi_n(\lambda)$ equals 0 when 
$\lambda=1$, $\tanh(\beta)=1$ and $n \to 0$. 

To find the proper maximum and perform the $n \to 0$ limit
is much more difficult when the temperature is positive i.e. $\tanh(\beta) < 1 $.
The problem is that we were unable to find
an analytic expression for any $n$ of the sum
$\sum_{\sigma} x(\sigma) \log( x(\sigma))$ when $0 <\lambda < 1$.
Nevertheless, in order to find the behavior of the free energy in the 
paramagnetic region  where $\lambda=0$ 
and just around the transition  where $\lambda \ll 1$ it is sufficient to
expand $\Phi_n$ with respect to $\lambda$.
The minimum order of the expansion is the fourth, in fact,
lower orders would be unable to provide a maximum. 

Let us rewrite $x(\sigma)= \frac{1}{2^n}(1+y(\sigma))$ where 
$y(\sigma)= \theta_c (\prod_{\alpha=1}^n (1+\lambda \, \sigma^\alpha)-1)$,
then notice that $y(\sigma)$ is a polynomial of $\lambda$ with 
no terms of order 0. Then we can expand  $\sum_{\sigma} x(\sigma) \log( x(\sigma))$
to the fourth order in $y(\sigma)$, the approximate expression 
can be easily computed in a compact analytic form in terms of $n$ and $\lambda$
and coincides with  $\sum_{\sigma} x(\sigma) \log( x(\sigma))$
for all terms of order 4 or less of the expansion in $\lambda$.

This analytic approximation can be inserted in (\ref{phitemp}) 
replacing   $\sum_{\sigma} x(\sigma) \log( x(\sigma))$ and we obtain 
an expression 
whose expansion coincides with that of $\Phi_n(\lambda)$ 
up to terms of order 4 in $\lambda$.
At this point, if we retain only these coinciding terms and
we expand to order 1 in $n$, we obtain 

\begin{equation}
\Phi_n (\lambda) \simeq n \theta_c^2 \left [\frac{\lambda^2}{2}(\gamma \tanh(\beta)-1)
- \frac{\lambda^4}{12}(3\gamma \tanh(\beta)^2-3 +6 \theta_c -2\theta_c^2)
 \right]+ n\log(2)
\label{philambda}
\end{equation}

The maximum is reached in $\lambda=0$ for $\gamma \tanh(\beta) \le 1$
and in $\lambda=\lambda_c$ given by

\begin{equation}
\lambda_c^2 = \frac{\gamma \tanh(\beta)-1}{\gamma \tanh(\beta)^2-1 + 2\theta_c
-\frac{2}{3}\theta_c^2}
\label{lambda}
\end{equation}
when $\gamma \tanh(\beta) > 1$.
Notice that in this region $\gamma \tanh(\beta)^2 \ge \frac{1}{\gamma}$,
therefore the denominator in (\ref{lambda})  is larger then 
$\frac{1-\gamma}{\gamma} + 2\theta_c-\frac{2}{3}\theta_c^2$ which, in turn,
is larger then 0.

The conclusion is that the magnetization vanishes when $\gamma \tanh(\beta) \le 1$
and equals  $m_c=\theta_c \lambda_c$ when $\gamma \tanh(\beta) > 1$ which implies
that we have exactly determined the transition line  $\gamma \tanh(\beta) =1$.

Inserting  (\ref{lambda}) in (\ref{philambda}) we obtain

\begin{equation}
- \beta F = \frac{\gamma}{2} \log(\cosh(\beta)) +\log(2)+ 
\frac{\theta_c^2}{4} \,
\frac{(\gamma \tanh(\beta)-1)^2}{\gamma \tanh(\beta)^2-1 + 2\theta_c
-\frac{2}{3}\theta_c^2}
\label{finallow}
\end{equation}
which gives the free energy
in the low (but not too low) temperature region $\tanh(\beta)>1$.
On the contrary, inserting $\lambda=0$ in (\ref{philambda}),
we obtain

\begin{equation}
- \beta F = \frac{\gamma}{2} \log(\cosh(\beta)) +\log(2)
\label{finalhigh}
\end{equation}
which gives the exact free energy
in the high temperature region $\tanh(\beta) \le 1$ where the 
magnetization vanishes. 

The free energy in the ferromagnetic region
can be computed at any necessary precision in a 
straightforward manner, it is sufficient in fact to expand  
$\sum_{\sigma} x(\sigma) \log( x(\sigma))$
to the higher orders in $\lambda$.
In fact, at any order, the expansion is a compact analytic expression 
of $\lambda$ and $n$.

Finally we remark that in the limit $\beta \to 0$, $\gamma \to \infty$ 
and $\gamma \beta= \tilde{\beta}$
one recovers the fully connected mean field Ising model.
In this limit, our expansion gives $\theta_c=1$, $m_c=\lambda_c$, 
$m_c^2=3(\tilde{\beta}-1)$ and 
$- \beta F= \frac{3}{4}(\tilde{\beta}-1)^2 + \log(2)$ which is the mean field
Ising model solution in the same approximation.

\section{6 - Discussion and outlook}

In this paper we obtained the exact solution of 
the dilute Ising Model in the case of a vanishing temperature 
even in presence of a magnetic field.
The solution gives a complete description of the geography
of the associated random graph, not only the 
size of the continent but also
the number of isolated clusters of connected spins of all given 
size (islands) and the average number of bonds per spin both in the continent
and in the islands.
We were also able to find out the exact solution in the paramagnetic region
and exactly determine the phase transition line.
On the contrary, in the positive temperature ferromagnetic region, 
our solution (\ref{xtemp}) is the correct one only if
the magnetization $\lambda_c$ of the spins in the continent is self-averaging 
in the thermodynamic limit. 
This assumption is not proven and the possibility
that the magnetization has 
symmetry-broken characterization remains open,
nevertheless, since at zero temperature there is not breaking
and since the system is not frustrated,
we are quite confident that our choice is correct.
Furthermore, the ferromagnetic free energy for positive temperature
is only given in terms of an expansion with respect to the order parameter 
$\lambda$ and, although the expansion can be made as accurate as necessary, 
the full and explicit free energy is still to be computed, if this
is possible.

Finally, we would like to mention that the present strategy can be
straightforwardly applied to the dilute spin glass,
but, unfortunately, it seems much more
difficult to find analogous results.

\section{Acknowledgments}

We are deeply grateful to Michele Pasquini for advices, suggestions and
a critical reading of the manuscript which permitted many improvements.
In particular, he provided results contained in (\ref{al}) and (\ref{sal}).
We also thank Daniel Gandolfo, Armando G. M. Neves
and Filippo Petroni for many discussions.

\end{document}